\documentclass[aps,prl,preprint,superscriptaddress]{revtex4}

\usepackage{graphicx}% Include figure files
\usepackage{dcolumn}% Align table columns on decimal point
\usepackage{bm}% bold math

\begin{document}

%\preprint{}

\title{Intrinsic Quantum Correlations of Weak Coherent States for Quantum Communication.}


\author{Yong Meng Sua, Erin Scanlon, Travis Beaulieu, Viktor Bollen, and Kim Fook Lee}
\email[]{kflee@mtu.edu}
\affiliation{%
Department of Physics,\\ Michigan Technological
University,\\ Houghton, Michigan 49931}%
%\thanks{ }
%\affiliation{}

% Collaboration name if desired (requires use of superscriptaddress
% option in \documentclass). \noaffiliation is required (may also be
% used with the \author command).

\date{\today}

\begin{abstract}
Intrinsic quantum correlations of weak coherent states are observed
between two parties through a novel detection scheme, which can be
used as a supplement to the existence decoy-state BB84 and
differential phase-shift quantum key distribution (DPS-QKD)
protocols. In a proof-of-principle experiment, we generate
bi-partite correlations of weak coherent states using weak local
oscillator fields in two spatially separated balanced homodyne
detections. We employ nonlinearity of post-measurement method to
obtain the bi-partite correlations from two single-field
interferences at individual homodyne measurement.
 This scheme is then used to
demonstrate bits correlations between two parties over a distance of
10 km through a transmission fiber. We believe that the scheme can
add another physical layer of security to these protocols for
quantum key distribution.

\end{abstract}

%\keywords{}

\maketitle

%\section{}
%\subsection{}
%\subsubsection{}

Quantum entanglement and superposition provide secure communication
between two parties for key generation and information processing.
However, entanglement based key generation such as Ekert's
protocol~\cite{Ekert91} is hard to implement in real-world optical
fiber network because bi-partite correlations of entangled
photon-pairs are sensitive to loss. There has been much interest in
quantum key generation of using weak coherent states or highly
attenuated lasers. Quantum key distribution using weak coherent
states, such as coherent state differential phase-shift quantum key
distribution (DPS-QKD) ~\cite{Inoue02,Yamamoto09} and decoy-states
BB84 protocols~\cite{Lo05,Wang05,Lo06,Nam07,Peng07,Tobias07}, have
been proven to be unconditional secure against photon-number
splitting attack (PNS). The DPS-QKD uses intrinsic correlations
between the relative phase shifts $\{0,\pi\}$ of two consecutive
pulses to achieve unconditional security between two parties by
constructing equivalent states for the entanglement-based
protocol~\cite{Yamamoto09}. The decoy state quantum key distribution
uses intrinsic correlations between the relative mean photon numbers
of two set of weak coherent states to detect PNS attack in BB84
protocol~\cite{Lo05}. Meanwhile, Y00 protocol~\cite{Yuen03} uses
intrinsic correlations between phase and mean photon number
fluctuations of weak coherent states to provide cryptographic
service of data encryption between two parties.

Intrinsic quantum correlations of coherent states can be prepared,
measured and shared between two parties for quantum cryptography. We
are motivated to propose a scheme based on weak coherent states for
generating intrinsic bi-partite correlations as a supplement
resource to the existence protocols such as coherent state DPS-QKD
and decoy state BB84.

Weak local oscillator (LO) field in a coherent state has been
successfully used to directly measure bi-partite correlation
functions of a two-photon source and violate Bell's inequalities
using homodyne detection with photon counting~\cite{kuzmich01}. In
this work, we employ a weak local oscillator field in a coherent
state to extract intrinsic correlations of weak coherent states
between two parties using balanced homodyne measurement. Briefly, we
first prepare a weak coherent state using a highly attenuated laser
at telecom wavelength. The coherent state is split by a 50/50 beam
splitter and sent to Alice and Bob. Alice and Bob, each has a
balanced homodyne detection scheme for measuring his/her coherent
state with a weak local oscillator field. We employ nonlinearity of
post-measurement method, i.e., multiply two single-field
interferences from individual balanced homodyne measurement. Then,
the mean-value of the multiplied signal provides raw data of
field-field correlations of weak coherent states. We normalize the
raw data with the mean photon numbers of weak coherent state and LO
field to obtain the coherent state bi-partite correlation function
(CSBC). It is named so as to avoid confusion with bi-partite
correlations provided by four Bells states of polarization-entangled
photon pair. Four types of correlation functions $\pm\cos
2(\theta_{1}\pm\theta_{2})$ can be prepared by using linear optics
devices in Alice or Bob alone, where $\theta_{1}$ and $\theta_{2}$
are the projection angles of the analyzers at Alice and Bob. This
means that Alice can keep her copy of the coherent state and send
another copy to Bob. By locally changing the relative phases between
her coherent state and weak local oscillator field, her acts will
change the correlation functions shared with Bob. Once we establish
one of the four correlation functions between Alice and Bob over a
distance of 10 km through a transmission fiber, we change the phases
of the weak local oscillator field $\{0,\pi\}$ for implementing bits
correlations between them.

The weak coherent state $|\alpha\rangle$ and weak LO field in a
coherent state $|\beta\rangle$ are treated as a product state of two
independent coherent states $|\alpha, \beta\rangle$ in the input of
the beam splitter~\cite{Ou87}. The density matrix of the output
state from the beam splitter is depended on the integration over the
phase space of P-representation for the input product state. Since
the $|\alpha\rangle$ and $|\beta\rangle$ are intrinsically
correlated from the same laser through the LO phases $\{0,2\pi\}$,
then integration over the phase-spaces of $\alpha$ and $\beta$ can
produce the output state that is intrinsically
entangled~\cite{Schleich01}. This is accomplished in our experiment
by conducting the mean-value measurement of the multiplied signals
of the output state. The two mode coherent states at the output of
the beam splitter, $|\alpha +
\beta\rangle_{1}|\alpha-\beta\rangle_{2}$, can be manipulated by
linear phase shifters to project out coherent interferences parts
$|\alpha_{1} \beta_{2}\rangle \pm |\beta_{1}\alpha_{2}\rangle$ or
$|\alpha_{1} \alpha_{2}\rangle \pm |\beta_{1}\beta_{2}\rangle$ plus
phase-space noises of $|\alpha\rangle$ and $|\beta\rangle$.

\begin{figure}
\includegraphics[scale=0.5]{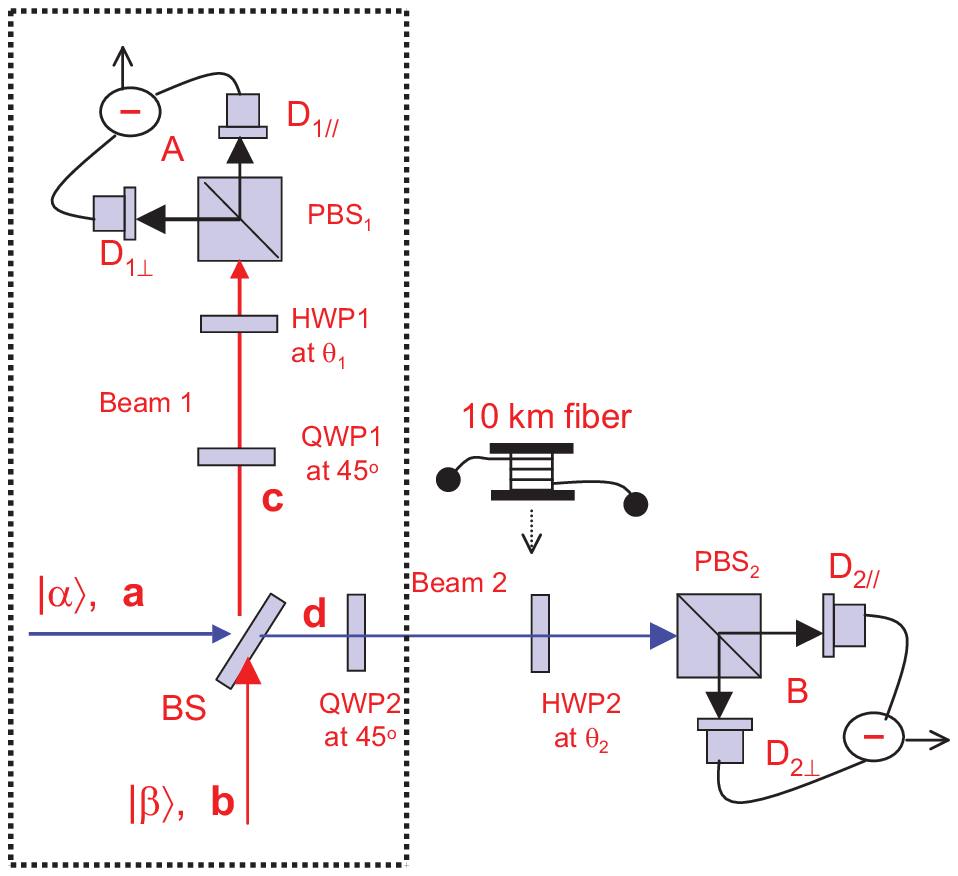}
\caption{Experiment setup for observing coherent state bi-partite
correlation of weak coherent states.}
\end{figure}

A proof-of-principle experiment for the brief description above is
shown in Fig.1. A highly attenuated laser at telecom-band wavelength
of 1534 nm is used to provide weak coherent states and weak local
oscillator field. We use a 50/50 beam splitter (BS) to mix the
horizontally polarized coherent state $|\alpha\rangle$ and the
vertically polarized local oscillator field $|\beta\rangle$. The
input field operators $\hat{a}$ and $\hat{b}$ at the beam splitter
are the annihilation operators for the coherent state
$|\alpha\rangle$ and the LO field $|\beta\rangle$, respectively.
Using unitary transformation matrix of a 50/50 beam splitter, the
output modes at the beam splitter are
$\hat{c_{1}}=\frac{1}{\sqrt{2}}(\hat{a} \textbf{x})+
i\hat{b}\textbf{y}$ in beam 1 and
$\hat{d_{2}}=\frac{1}{\sqrt{2}}(i\hat{a} \textbf{x})+
\hat{b}\textbf{y}$ in beam 2. A quarter wave plate (QWP) at
$45^{\circ}$ is inserted at beam 1 and beam 2 to transform the
linearly polarized states to circularly polarized states. Then,
after the matrix transformation of a quarter wave plate, the field
operators $\hat{c_{1}}\rightarrow
\hat{c_{1}}'=\frac{1}{\sqrt{2}}((\hat{a}+\hat{b})\textbf{x}+
i(\hat{b}-\hat{a})\textbf{y})$ and $\hat{d_{1}}\rightarrow
\hat{d_{1}}'=\frac{1}{\sqrt{2}}((i\hat{a}-\hat{b})\textbf{x}+
(\hat{b}+\hat{a})\textbf{y})$. A half-wave plate HWP1 (HWP2) is
inserted in beam 1 (2) before a cube polarization beam splitter PBS1
(PBS2) to project out the polarization state
$\theta_{1}$($\theta_{2}$) with unit vector $\hat{e_{1(2)}}=\cos
\theta_{1(2)}\textbf{x}+\sin \theta_{1(2)}\textbf{y}$, respectively.
The field operators
$\hat{c_{1}}''=(\hat{c_{1}}'\cdot\hat{e_{1}})\hat{e_{1}}$ after the
PBS1 at beam 1 and
$\hat{d_{2}}''=(\hat{d_{2}}'\cdot\hat{e_{2}})\hat{e_{2}}$ after the
PBS2 at beam 2 provide the photon number operators of the
transmitted component of the combined fields as,

\begin{equation}
\hat{c^{*}_{1}}''\hat{c_{1}}''=\frac{1}{2}[\hat{a^{*}\hat{a}}+\hat{b^{*}}\hat{b}
+\hat{a^{*}}\hat{b}e^{i2\theta_{1}}+\hat{b^{*}}\hat{a}e^{-i2\theta_{1}}]\label{eq:01}
\end{equation}
\begin{equation}
\hat{d^{*}_{2}}''\hat{d_{2}}''=\frac{1}{2}[\hat{a^{*}\hat{a}}+\hat{b^{*}}\hat{b}
-\hat{a^{*}}\hat{b}e^{i2\theta_{2}}-\hat{b^{*}}\hat{a}e^{-i2\theta_{2}}].\label{eq:02}
\end{equation}
Now, using the input weak coherent state $|\alpha\rangle =|\alpha
e^{i\phi_{\alpha}}\rangle$ and the input weak LO field in a coherent
state $|\beta\rangle = |\beta e^{i\phi_{LO}}\rangle$ into
Eq.~\ref{eq:01} and Eq.~\ref{eq:02}, the detectors A and B measure
the transmitted components of the beat intensities,

\begin{equation}
I_{1\parallel}(\theta_{1}, \phi_{LO})\rightarrow\langle
\alpha,\beta|\hat{c^{*}}''\hat{c}''|\alpha,\beta\rangle=\eta_{A}[|\alpha|^{2}+|\beta|^{2}
+|\alpha||\beta|Cos(2\theta_{1}+\phi_{LO}-\phi_{\alpha})]\nonumber\\
\end{equation}
\begin{equation}
I_{2\parallel}(\theta_{2}, \phi_{LO})\rightarrow\langle
\alpha,\beta|\hat{d^{*}}''\hat{d}''|\alpha,\beta\rangle=\eta_{B}[|\alpha|^{2}+|\beta|^{2}
-|\alpha||\beta|Cos(2\theta_{2}+\phi_{LO}-\phi_{\alpha})]\\
\end{equation}
where $|\alpha, \beta\rangle$ is the input state of the beam
splitter and $\eta_{A}$($\eta_{B}$) is the conversion efficiency
(Watt $\rightarrow$ Current) for detection electronics of A (B). The
first two terms are intensities of the two coherent states and the
last term is the interference term consists of polarization angle
$\theta_{1}$ ($\theta_{2}$), the phases of LO ($\phi_{LO}$) and the
weak coherent state ($\phi_{\alpha}$). The beat intensity
$I_{1\parallel}(\theta_{1}, \phi_{LO} )$ is anti-correlated to
$I_{2\parallel}(\theta_{2}, \phi_{LO} )$ because of the $\pi$-phase
shift induced by the 50/50 beam splitter. The beat intensity for the
reflected signal at the PBS1 (PBS2)is $I_{1\perp}(\theta_{1\perp},
\phi_{LO})$ ($I_{2\perp}(\theta_{2\perp}, \phi_{LO})$), where the
$\theta_{1\perp}=\theta_{1}+\pi/2$
($\theta_{2\perp}=\theta_{2}+\pi/2$), respectively. Then, the
detectors A and B measure the balanced homodyne beat intensities,
that are,
\begin{equation}
\mathcal{A}_{1} \rightarrow I_{1\parallel}-I_{1\perp}= 2\eta_{A}
|\alpha||\beta|\cos
(2\theta_{1}+\phi_{LO}-\phi_{\alpha})\label{eq:05}
\end{equation}
\begin{equation}
\mathcal{B}_{2} \rightarrow I_{2\parallel}-I_{2\perp}= -2\eta_{B}
|\alpha||\beta|\cos
(2\theta_{2}+\phi_{LO}-\phi_{\alpha})\label{eq:06}
\end{equation}
plus shot noise at each detection of $\mathcal{A}_{1}$ and
$\mathcal{B}_{2}$. The individual intensities of the two coherent
states are subtracted. One can see that the information $2\theta_{1}
+ \phi_{LO}$ and $2\theta_{2}+\phi_{LO}$ are protected by quantum
phase noise $\phi_{\alpha}$ with phase fluctuation of $\Delta
\phi_{\alpha} \geq \frac{1}{\Delta n}$, where low mean photon number
fluctuation associated with phase fluctuation is provided by the
weak coherent state. The balanced homodyne beat intensities in
detectors A and B are then multiplied to obtain,
\begin{equation}
\mathcal{A}_{1} \mathcal{B}_{2}
%=-4\eta_{A}\eta_{B}|\alpha|^{2}|\beta|^{2}Cos(2\theta_{1}+\phi_{LO}-\phi_{\alpha})Cos(2\theta_{2}+\phi_{LO}-\phi_{\alpha})
=-2\eta_{A}\eta_{B}|\alpha|^{2}|\beta|^{2}[\cos
2(\theta_{1}-\theta_{2})+\cos
2(\theta_{1}+\theta_{2}+\phi_{LO}-\phi_{\alpha})].\label{eq:03}
\end{equation}
The multiplied balanced-homodyne beat intensities did not provide
bi-partite correlation function directly. This is predicted because
so far we have performed the multiplication of single-field
interferences obtained from individual balanced homodyne detection
at the detectors A and B. However, by taking the mean value of this
multiplied beat intensities, the last term is averaged to zero due
to slowly varying local oscillator phase $\phi_{LO}$ from $\{0, 2\pi
\}$ protected by randomness of quantum phase noise $\phi_{\alpha}$.
Note that $\Delta \phi_{\alpha}$ cannot randomly provide phase shift
$0\rightarrow 2 \pi$. And hence, we obtain the expectation value of
two detectors or coherent state bi-partite correlation function as
given by,
\begin{equation}
\overline{\mathcal{A}_{1} \mathcal{B}_{2}} \rightarrow
\langle\mathcal{A}_{1}\mathcal{B}_{2}\rangle = - 2
\eta_{A}\eta_{B}|\alpha|^{2}|\beta|^{2} \cos
2(\theta_{1}-\theta_{2}).\label{eq:08}
\end{equation}
The coherent state bi-partite correlation is protected by the term
$\cos 2(\theta_{1}+\theta_{2}+\phi_{LO}-\phi_{\alpha})$, which is
averaged to zero. In real practice, the product of mean photon
numbers $|\alpha|^2|\beta|^2$ is obtained by setting the correlation
function to its maximum obtainable value, that is,
$\theta_{1}=\theta_{2}$. The raw data of the multiplied beat
intensities is then normalized with the product of
$2\eta_{A}\eta_{B}|\alpha|^2|\beta|^2$ to obtain correlation
function $-\cos 2(\theta_{1}-\theta_{2})$. As for the use of quantum
communication between two distant observers, we have to establish
the bi-partite correlation function and then to implement bits
correlations at detectors A and B through the LO phases $\{0,\pi\}$
without using the post-measurement method.
\begin{figure}
\includegraphics[scale=0.4]{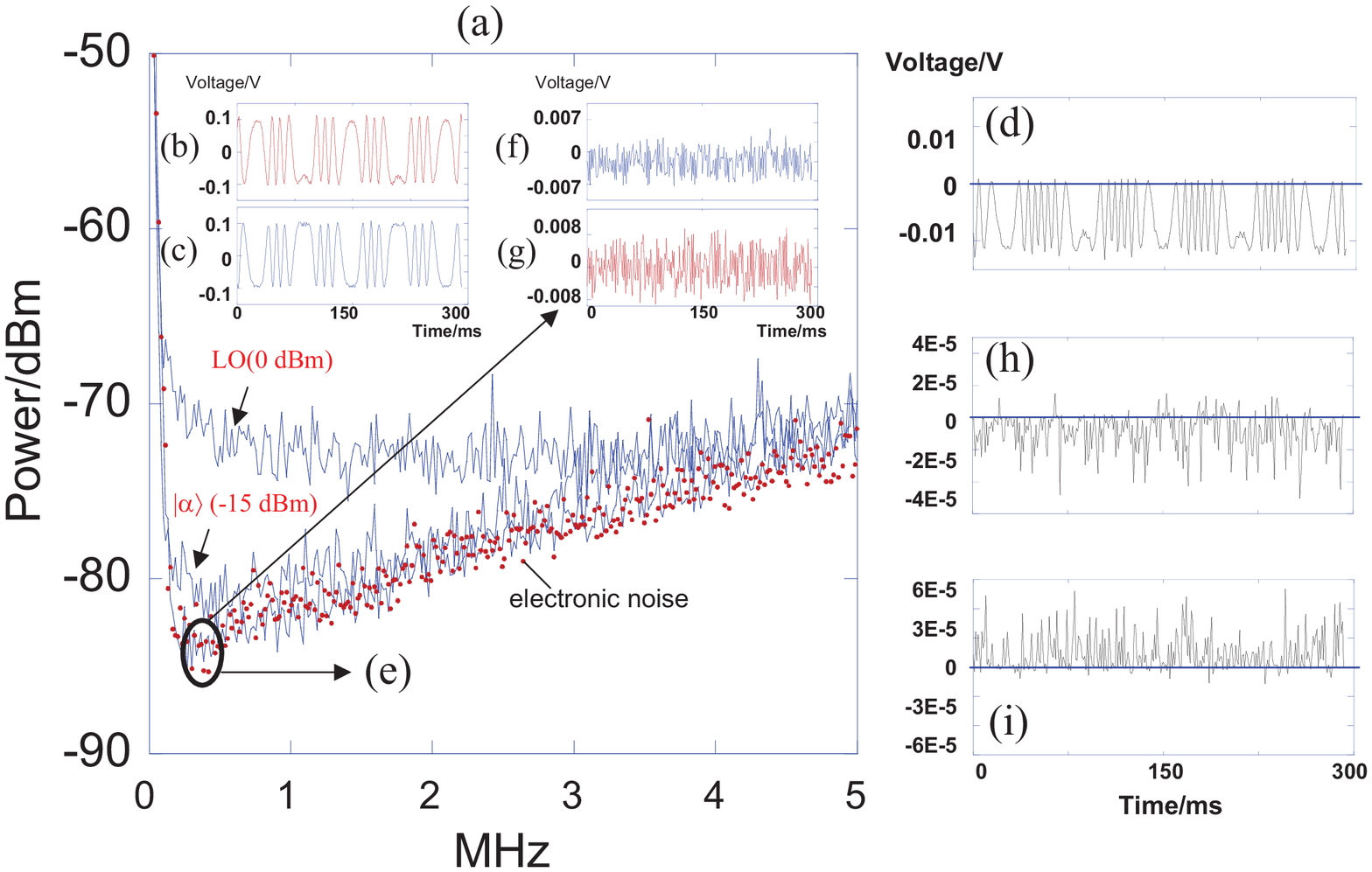}
\caption{(a) Shot noises of the weak LO field (0 dBm) and coherent
state (-15 dBm), and the corresponding beat signals for
$\theta_{1}=\theta_{2}$ at detector A (b) and detector B (c), and
their multiplied signal (d). (e) shot noises for the weak LO field
(-30 dBm) and the corresponding beat signals for
$\theta_{1}=\theta_{2}$  at detector A (f) and detector B (g), and
their multiplied beat signal (h). (i) the multiplied beat signal
when the relative angle $\theta_{1}-\theta_{2}=\pi/2$. (square dot)
Electronic noise.}
\end{figure}
To verify the analysis above, we perform systematic studies of the
experiment. The balanced homodyne detectors are made of two PIN
photodiodes (EXT500). We use a piezoelectric transducer (PZT) to
ramp the phase of the weak LO beam. We first perform the experiment
by using a strong LO field and a coherent state with average power
of 0 dBm and -15 dBm, respectively. Figure 2 (a) shows the spectrum
of the shot noise levels of the strong LO field and coherent state,
and the electronic noise of our detection system. We set the
relative angle between Alice's and Bob's analyzers as
$\theta_{1}-\theta_{2}=0$. The beat intensities at detectors A and B
are shown in Fig. 2(b) and (c)(inset of Fig.2(a)). The discontinuity
of the beats are due to the ramping of the PZT. With these large
mean photon fluxes, the interference signals are stable as predicted
by the coherent states with large mean photon number. The product of
the beat intensities is shown in Fig.2(d), which is a clear
indication of anti-correlation between single-field interferences at
detectors A and B. Then, we attenuate the laser light to obtain weak
LO field and weak coherent state with average power of -30 dBm and
-30 dBm, respectively. All the average optical powers reported in
this work are measured just before the PBS1(PBS2). Figure 2(e) shows
the shot noise of weak LO field almost falls on the electronic noise
spectrum. We observe the beat intensities at detectors A and B as
shown in Fig. 2(f) and (g) (inset of Fig.2(a)) with the interference
signals hidden or protected by the shot noises of the LO field,
quantum phase noise $\phi_{\alpha}$ of the weak coherent state due
to low mean photon number fluctuation, and electronic noises. These
are predicted by the Eq.~\ref{eq:05} and Eq.~\ref{eq:06}. In the
experiment, the beat intensities at detectors A and B are stored in
their computers. Then, these raw data are multiplied together as
shown in Fig. 2(h). The multiplied beat intensity consists of two
parts; coherent and noise interferences. The coherent interference
part contains the term $- 2\eta_{A}\eta_{B} |\alpha|^{2}|\beta|^{2}
\cos 2(\theta_{1}-\theta_{2})$. The noise interference part contains
the term $- 2 \eta_{A}\eta_{B}|\alpha|^{2}|\beta|^{2} \cos
 2(\theta_{1}+\theta_{2}+\phi_{LO}-\phi_{\alpha})$ which is averaged
to zero because of the periodic of the LO phase, $\phi_{LO} = \{0,
2\pi\}$, protected by the quantum phase noise $\phi_{\alpha}$. The
contribution of short noise is averaged to zero. Since the
electronic noise is not completely random and present in our
measurement method, the noise will create statistical errors in the
mean-value measurement of the coherent part of the multiplied
signal. Note that the measurement method is also applied for large
mean photon number coherent states as shown in Fig.2(d) and also for
the mixture of stable and noise fields~\cite{Lee09}. The aim of the
post-measurement method is to make sure that weak coherent state for
quantum key distribution can provide coherent state bi-partite
correlation, which can be used as a supplement for the decoy state
BB84 and coherent state DPS-QKD. From the Fig.2(h), we obtain the
product of the mean photon number fluxes for the
$|\alpha|^2|\beta|^2$, where the $-\cos
 2(\theta_{1}-\theta_{2})=-1$ is maximum obtainable value for
$\theta_{1} =\theta_{2}$. Fig. 2(i) shows that the multiplied signal
is proportional to  $-\cos
 2(\theta_{1}-\theta_{2})=1$ when the relative angle is set to
$\theta_{1}-\theta_{2}=\pi/2$. We are able to prepare four types of
bi-partite correlations such as $\pm \cos 2(\theta_{1}\pm
\theta_{2})$ shared between two parties by using liner phase
shifters on either beam 1 or beam 2. For practical quantum
communication, Alice can keep the beam 1 and linear phase shifters
as highlighted in the box in Fig.1, and send out the beam 2 to Bob.
Since Alice can change the phases of beam 1 locally, her acts will
change the coherent state bi-partite correlation function shared
with Bob. Fig.3(a) shows that the normalized coherent state
bi-partite correlation function $-\cos 2(\theta_{1}-\theta_{2})$.
For each data point, we take 10 shots of the multiplied signal and
obtain the average mean-value. The error bar is mainly due to the
electronic noises. The offset of the relative angle due to
imperfection of quarter wave-plates has been corrected.
\begin{figure}
\includegraphics[scale=0.4]{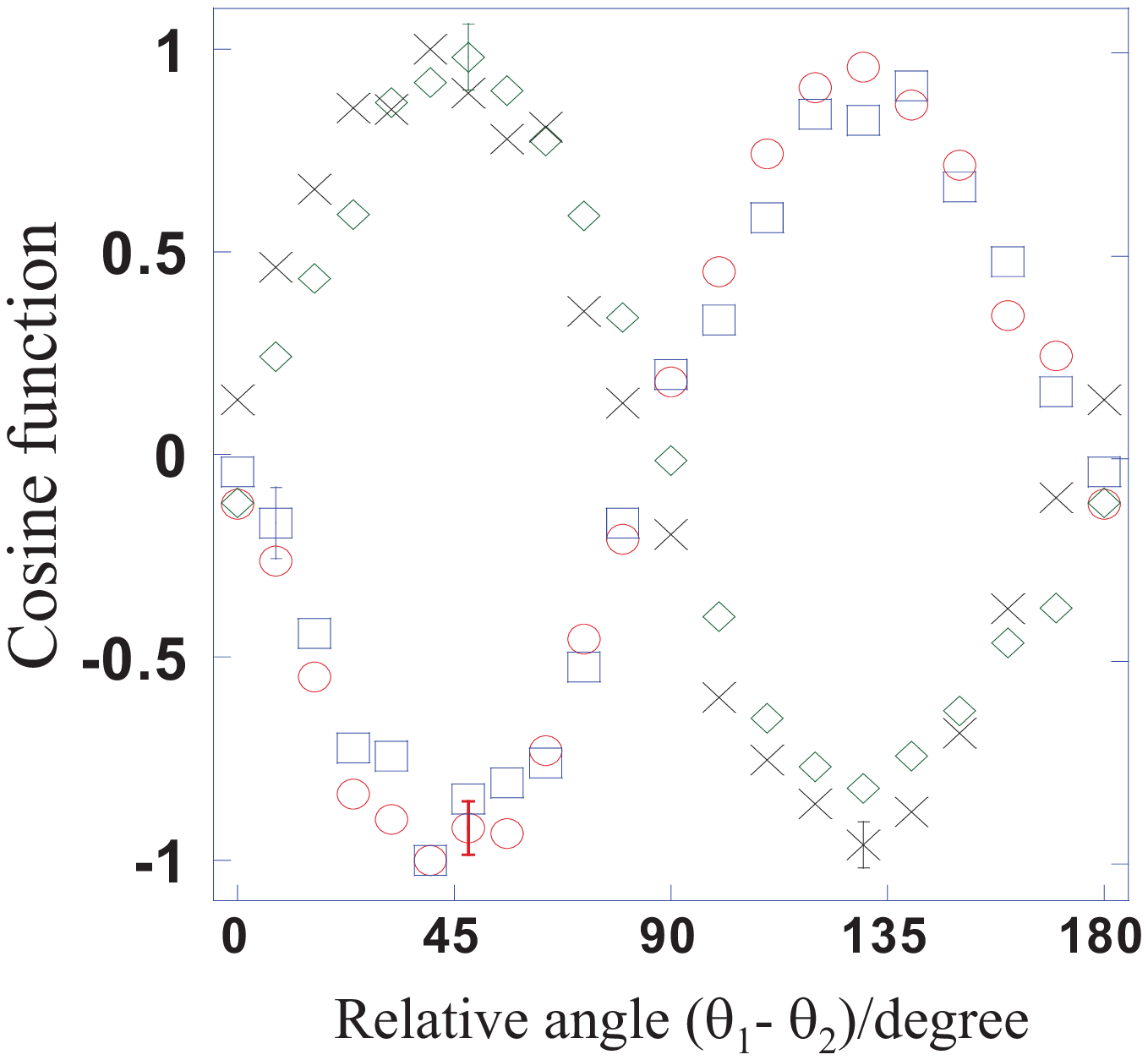}
\caption{The coherent state bi-partite correlations for
(a)(box)$-\cos 2(\theta_{1}-\theta_{2})$, (b)(diamond)$\cos
2(\theta_{1}+\theta_{2})$, (c)(cross)$\cos2(\theta_{1}-\theta_{2})$,
and (d)(circle) $-\cos 2(\theta_{1}+\theta_{2})$.}
\end{figure}
For preparing the correlation function of $\cos
2(\theta_{1}+\theta_{2})$ in Fig.3(b), we set the $\lambda/4$
wave-plate in beam 1 to -$45^{\circ}$ so that the beat intensity
$\mathcal{A}_{1}$ of Eq.~\ref{eq:05} becomes $\mathcal{A}_{1}
\propto -\cos (2\theta_{1}-(\phi_{LO}-\phi_{\alpha}))$. As for the
correlation function of $\cos 2(\theta_{1}-\theta_{2})$ in Fig.
3(c), we insert a $\lambda/2$ plate at $0^{\circ}$ in beam 1 so that
the minus sign of beat intensity $\mathcal{A}_{1}$ of
Eq.~\ref{eq:05} is changed to positive sign. Similarly, with the
$\lambda/2$ wave-plate at $0^{\circ}$ and the $\lambda/4$ wave-plate
at -$45^{\circ}$ in beam 1, the beat signal $\mathcal{A}_{1}$ of
Eq.~\ref{eq:05} becomes $\cos
(2\theta_{1}-(\phi_{LO}-\phi_{\alpha}))$. Thus, the correlation
function of $-\cos 2(\theta_{1}+\theta_{2})$ is obtained as shown in
Fig.3(d).

After we establish or choose one of coherent state bi-partite
correlation functions between Alice and Bob, we implement bits
correlations between them. To perform this measurement for the
established correlation function of $-\cos
2(\theta_{1}-\theta_{2})$, we ramp the PZT to obtain one period of
interference signal. We reduce the average power of the weak LO
field to -39 dBm and the average power of the weak coherent state to
-39 dBm. The output of the balanced homodyne beat intensity at the
detector A is directly connected to a lock-in-amplifier, where the
reference frequency at 168 Hz is obtained from a function generator
that drives the PZT. We measure quadrature phases of weak coherent
state with the step size of $n\pi/2$ (n=integer) as shown in
Fig.4(a)(solid line). Using the same lock-in reference phase in the
lock-in-amplifier, we measure the quadrature phases of weak coherent
state at detector B as shown in Fig.4(a)(dashed-line).  We have
observed the bits correlations between two parties for the shared
correlation function of $-\cos 2(\theta_{1}-\theta_{2})$ as shown in
Fig.4 (a), where the positive (negative) quadrature signal is
encoded as keys/bits '1' ('0'), respectively. By using the same
lock-in reference phases, we observe bits correlations for other
three types of correlation functions $-\cos 2
(\theta_{1}+\theta_{2})$, $\cos 2(\theta_{1}+\theta_{2})$, and $\cos
2(\theta_{1}-\theta_{2})$ as shown in Fig.4(b), (c) and (d),
respectively. In real practice, we can establish one of the CSBC for
calibrating the lock-in reference phases at Alice and Bob.
\begin{figure}
\includegraphics[scale=0.4]{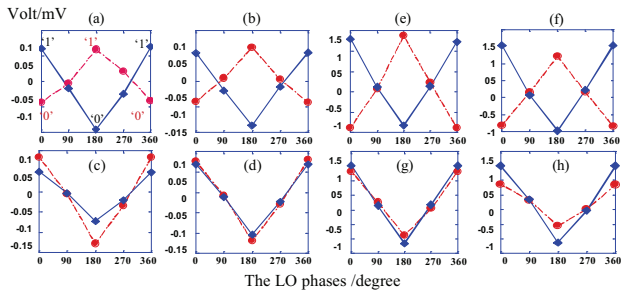}
\caption{Bits correlations for CSBC of ((a)and (e))$-\cos
2(\theta_{1}-\theta_{2})$;((b)and(f))$-\cos
2(\theta_{1}+\theta_{2})$;((c)and (g)) $\cos
2(\theta_{1}+\theta_{2})$;((d) and (h))$\cos
2(\theta_{1}-\theta{2})$ , where (e), (f), (g) and (h) are with the
10 km fiber. The size of the data point is the error bar for the
measurement.}
\end{figure}

We test the scheme by performing bits correlations between two
parties over a distance of 10 km through a transmission fiber. We
couple the beam 2 into the transmission fiber. A quarter wave plate
and a half wave plate (not shown in Fig.1) are used at the output of
the transmission fiber to compensate the birefringence. Since there
are losses in the couplings and the fiber, we use weak coherent
state with average power of -33 dBm and weak local oscillator field
with average power of -33 dBm before the homodyne detections. We
have established four types of correlation functions and performed
bits correlations for each shared correlation function between two
parties as shown in Fig.4(e), (f), (g) and (h).

As a supplement to the DPF-QKG, the phase of the weak LO field can
be randomly modulated as $\{0, \pi\}$ at certain frequency. Then,
the bits/keys correlations can be realized based on the established
CSBC shared by both parties. Since the established CSBC is
normalized with the product of mean photon numbers
$|\alpha|^{2}|\beta|^{2}$, photon number splitting attack can be
detected by adding a weak LO beam in the decoy state BB84 protocol
to check the CSBC shared between two parties. The security analysis
of the scheme is out of the scope of this paper.

Intrinsic correlations of coherent light field have been utilized to
implement entanglement~\cite{Lee04}, Grover search
algorithm~\cite{Lloyd00,Spreeuw01}, quantum
lithography~\cite{Hemmer06}, factoring number~\cite{Bigourd08} and
quantum walk~\cite{Perets08} through different well-designed
interference measurement methods. Intrinsic correlations of coherent
states do not exhibit non-locality as two-photon source. The
realization of intrinsic quantum correlation of weak coherent state
by using the measurement method is a first step toward linear-optics
quantum computing with weak light fields and single-photon source.

\begin{acknowledgments}
The authors would like to acknowledge that this paper is prepared
under the support of the start-up fund from Department of Physics,
Michigan Technological University. Erin Scanlon and Viktor Bollen
would like to acknowledge that the support of SURF to carry out the
experiment under supervision of Kim Fook Lee.

\end{acknowledgments}

%\bibliographystyle{apsrev}
%\bibliography{thesis}
% pasted in .bbl file dsf-pol-ent-pk.bbl

\newcommand{\noopsort}[1]{} \newcommand{\printfirst}[2]{#1}
  \newcommand{\singleletter}[1]{#1} \newcommand{\switchargs}[2]{#2#1}

\newpage

\pagebreak

\newpage

%\setcounter{figure}{0}
%\begin{figure}
%\begin{center}\
%\includegraphics[scale=0.5]{}

%\end{center}
%\caption{}
%\end{figure}

%\begin{figure}
%\begin{center}\
%\includegraphics[scale=0.5]{}
%\end{center}
%\caption{}
%\end{figure}

\end{document}